\documentclass[prb,superscriptaddress,twocolumn,amsmath,amssymb,floatfix]{revtex4-1}
\usepackage{latexsym}
\usepackage{dcolumn}
\usepackage[dvips]{graphicx}
\usepackage{amssymb}
\usepackage{graphicx}
\usepackage{psfrag, color}
\usepackage{verbatim}
\usepackage{amsmath}
\usepackage{epsf}
\usepackage{bigstrut}
\usepackage{soul}

\newcommand{\vk}{{\bf k}}

\begin{document}


\title{Charge transport in gapless electron-hole systems with arbitrary band dispersion}

\author{S. Das Sarma}
\affiliation{Condensed Matter Theory Center, Department of Physics, 
	 University of Maryland,	 College Park, Maryland  20742-4111}
\affiliation{Joint Quantum Institute, University of Maryland, College Park, Maryland 20742, USA}	 
\author{E. H. Hwang}
\affiliation{Condensed Matter Theory Center, Department of Physics, 
	 University of Maryland,	 College Park, Maryland  20742-4111}
\affiliation{SKKU Advanced Institute of Nanotechnology and Department
  of Physics, Sungkyunkwan University, Suwon 440-746, Korea } 

\date{\today}

\begin{abstract}
Using the semiclassical Boltzmann transport theory, we analytically consider dc charge transport in gapless electron-hole (both chiral and non-chiral) systems in the presence of resistive scattering due to static disorder arising from random quenched impurities in the background. We obtain the dependence of the Boltzmann conductivity on carrier density and temperature for arbitrary band dispersion in arbitrary dimensionality assuming long-range ($\sim 1/r$) Coulomb disorder and zero-range white noise disorder [$\sim \delta(r)$]. We establish that the temperature and the density dependence of the Boltzmann conductivity manifests scaling behaviors determining respectively the intrinsic semimetallic or the extrinsic metallic property of the gapless system.
Our results apply equally well to both chiral and non-chiral gapless systems, and provide a qualitative understanding of the dependence of the Boltzmann conductivity on the band dispersion in arbitrary dimensionality.
\end{abstract}

\maketitle

\section{introduction and background}
Much interest has focused over the last 10 years on the electrical transport properties of electron (and hole) systems where the band dispersion deviates from the standard $E(\vk) = \hbar^2 k^2/2m$ parabolic metallic band dispersion ($E$, $\vk$, $m$ respectively being single-particle energy, wave vector, and effective mass). The most important example of such a system in nature, which has been studied extensively over the last 10 years \cite{dassarma2011,peres}, is graphene\cite{geim2004} which has two-dimensional gapless linear relativistic Dirac-like chiral energy dispersion defined by $E(\vk)=\hbar v_F k$, with $k=|\vk|$ as the 2D wave number and $v_F$ as the graphene velocity defining its linear dispersion.

Very recent interest in the literature \cite{zyuzin2012} has been on 3D Dirac-Weyl materials which have chiral relativistic linear gapless band dispersion in three dimensions (i.e., akin to `3D graphene'). For our purpose the distinction between (degenerate) Dirac and (non-degenerate) Weyl systems is unimportant, and we will refer to the linearly dispersing gapless chiral systems as Dirac materials (with a certain degeneracy factor `$g$', which counts the number of carrier `species' or `flavors' in the system as reflected in the spin and/or valley degeneracy -- e.g., for graphene $g=4$ and for some of the proposed 3D Dirac materials $g=24$), completely ignoring any inter-species scattering (e.g., scattering between valleys in the band structure) and assuming each carrier species to be contributing independently (and equally) to charge transport. 
(The neglect of inter-species (often, inter-valley) scattering is an excellent approximation in most experimental situations since such scattering involves very large momentum transfer.)
A number of recent theoretical publications in the literature study transport properties in 3D Dirac systems \cite{dassarma2015,skinner2014,koshino2014,koshino2015,brouwer2014,biswas2014}. The number of publications on 2D graphene is huge ($>10,000$) and we can only cite some relevant review articles \cite{dassarma2011,peres}.

Our current work presented in this paper asks the following question: What can we say theoretically about transport properties of a gapless electron-hole system with arbitrary energy band dispersion (and in arbitrary dimensions)? Of course, the general theory for such a completely general problem is intractable, and we focus on the semiclassical transport properties within the Boltzmann theory using the relaxation time approximation. We also restrict ourselves to the dc electrical conductivity limited only by disorder scattering where the disorder arises from random quenched impurities in the material with electron-impurity scattering potential being either the long-ranged Coulomb potential (assuming to be the usual $1/r$-type in all dimensions where $r$ is the electron-impurity spatial separation) or the zero-range $\delta$-function potential. We consider both screened and unscreened limits of the Coulomb disorder, and obtain completely analytical results for the dc charge conductivity in all dimensions with our results agreeing with the known results in two and three dimensional gapless Dirac systems.
It turns out that our analytical results obtained in this work are very general, and apply to both chiral and non-chiral electron-hole gapless systems although most existing gapless systems (e.g. graphene) happen to be chiral in nature, obeying some effective form of massless Dirac-Weyl equations at long wavelength.

Our motivation for our theoretical work is obvious. Given the great current interest in gapless linearly dispersing Dirac systems, it makes sense to obtain results for both linear and quadratic energy dispersions in arbitrary dimensions in order to develop intuition about the dispersion-dependence of transport properties. Note that we restrict the system dimensionality to be greater than 1 (i.e., $d>1$, where $d$ is the system dimensionality) throughout this paper. 
If theoretical results can actually be obtained in arbitrary dimensions ($d>1$) for arbitrary band dispersions, obviously that would provide a more complete picture for the qualitative nature of transport properties in electronic materials, and doing this is therefore our primary motivation.
But there is actually a second subtle (indirect) reason arising from quantum criticality for 
our theoretical work, which we discuss below.

It has been known for almost 30 years, since the seminal work of Fradkin\cite{fradkin1986}, that a system of gapless electrons, where conduction (electron) and valence (hole) bands touch at a finite number of isolated points (so-called Dirac points where the system is a perfect semimetal with no free electrons or holes at zero temperature) in momentum space manifesting linear band dispersion, has a disorder-induced $T=0$ quantum phase transition for $d>2$ 
(with $d_c=2$ being the critical dimension). The transition manifests itself as the density of states (DOS), $D(E)$, becoming finite at $E=0$ (taking the Dirac point to be at $E=0$ with no loss of generality as our notation for zero energy) for infinitesimal disorder strength (finite critical disorder strength) for $d=2$ ($d>2$). For linearly dispersing carriers, of course, $D(E)$ goes as $E$ ($E^2$) in $d=2$ ($d=3$) in a clean noninteracting system with $D(E=0)=0$ at the Dirac point in both cases. The DOS serves as the nominal order parameter for this quantum phase transition, being zero (finite) for disorder strength below (above) the critical point (with the critical disorder being zero for $d=2$). Although the original result was obtained within an $1/N$-expansion with $N=g$ being the number of fermion species (or the electron degeneracy factor), this disorder-driven quantum phase transition has now been verified using many disparate theoretical techniques including one-loop\cite{goswami2011} and two-loop\cite{roy2014} renormalization group (RG) calculations, exact numerical simulations \cite{brouwer2014,herbut2014,pixley2015}, self-consistent Born approximation calculations\cite{koshino2014,koshino2015}, and general scaling theories \cite{gurarie2015}. In fact, very recent work\cite{roy2015} indicates that this disorder-induced quantum phase transition even survives 
some finite interaction effects leading to rich multicritical  behavior. 

We emphasize five important caveats with respect to this quantum phase transition: (i) it occurs in the undoped intrinsic situation when the chemical potential is precisely at the Dirac point ($E=0$) with no free electrons/holes present in the system (indeed for any finite doping, the chemical potential would shift from the Dirac point into the conduction or the valence band leading to $D(E=E_F)$ at the fermi level to be finite any way even in the clean system indicating the system to be a 'metal' with finite DOS); 
(ii) the transition has only been theoretically established for short-ranged (e.g., Gaussian) disorder; (iii) the so-called Griffiths physics associated with rare disorder regions may affect the transition leading to band tailing corrections even at $E=0$ \cite{nand2014} which could lead to finite DOS for all disorder in all dimensions (thus rendering the transition into a crossover); 
(iv) the physics of disorder-induced electron-hole puddle formation \cite{skinner2014} which is known to be of great importance \cite{adam2007,li2012}and relevance \cite{chen2008,martin2008} in graphene, is ignored (indeed disorder induced puddles produce to macroscopic density inhomogeneity in the system with considerable spatial fluctuations in the chemical potential which lead to the local DOS being always a fluctuating finite quantity throughout the system for any finite disorder even when the average Fermi level is at $E=0$; (v) all quantum localization effects are ignored consistent with the assumption of no intervally scattering in the system \cite{aleiner2006}.

The disorder-induced `semimetal-to-metal' transition (with the DOS increasing from zero in the semimetal to finite in the metal
at the Dirac point) discussed above provides an important (albeit indirect) context for our current work. The Boltzmann conductivity at $T=0$ for an electron system is given by $\sigma \propto D(E_F) \tau(E_F)$, and obviously the DOS plays a decisive role in directly determining the conductivity through the $D(E_F)$ factor as well as indirectly through the relaxation time $\tau$ which itself, of course, depends on the DOS (and also on the disorder potential). But, the DOS, $D(E)$, for an arbitrary band dispersion defined by $E(k) \sim k^{\alpha}$ goes as $D \sim E^{d/\alpha-1}$ in $d$-dimensions (e.g., for linear dispersion, $\alpha=1$, $D \sim E$, $E^2$ in $d=2$, 3, and for parabolic dispersion, $\alpha=2$, $D\sim E^0$, $E^{1/2}$ in $d=2$, 3, respectively, as is well-known), and thus specific combinations of ($\alpha,d$) provide finite or vanishing DOS at $E=0$. Thus, for $d=2=\alpha$ (i.e., the ordinary parabolic metallic electron system), the DOS is an energy-independent constant which is finite at $E=0$ whereas for $d=3=\alpha$ the same is true in 3D systems. 
In fact, for any $d=\alpha$ ($d>\alpha$) the DOS is finite (zero) at $E=0$ in arbitrary dimensions for the clean noninteracting system, thus making $\alpha =2$ (3) special for $d=2$ (3) respectively in the consideration of the putative semimetal-metal theoretical consideration.  We note that the critical dimensionality for the disorder-driven (through the DOS) semimetal-metal transition is precisely $d_c=2\alpha$ \cite{fradkin1986} showing the coupled role of $d$ and $\alpha$ in this context.
Thus, we  can compare the effect of having finite versus vanishing DOS on transport properties by tuning the band dispersion (i.e., $\alpha$) and the dimensionality (i.e., $d$). This should provide some qualitative insight into the behavior of the charge conductivity through the semimetal (``vanishing DOS") to metal (``finite DOS"), albeit by tuning $\alpha$ (rather than by disorder). We are also in a position to make statements about how the nature of disorder (long versus short range) affects the conductivity in the semimetallic versus the metallic phase.

Thus, the theoretical issue we address in the current paper is complementary to that considered by Fradkin and others in the context of disorder-driven quantum criticality \cite{koshino2014,koshino2015,brouwer2014,biswas2014,fradkin1986,goswami2011,roy2014,herbut2014,pixley2015,gurarie2015,roy2015,nand2014} -- we ask how the Boltzmann conductivity behaves as a function of the band dispersion parameter ($\alpha$) in different dimensions ($d>1$) for long- and short-range disorder assuming that the DOS of the gapless system is determined only by band dispersion (and dimensionality), and not by disorder (with disorder entering the calculated Boltzmann conductivity only through the transport relaxation time $\tau$).  Since the clean system DOS depends on $\alpha$ and $d$ (and may be finite or zero at the Dirac point, depending on the precise values of $\alpha$ and $d$), we anticipate deriving general relations between the conductivity of the gapless system (within the Boltzmann theory where the DOS is not renormalized by disorder) and values of $\alpha$ and $d$ (which in turn determine whether the clean system is a semimetal with vanishing DOS or a metal with finite DOS). We are therefore establishing a connection between the clean system DOS and the semiclassical conductivity, and not exploring quantum critical physics.  
From the perspective of quantum criticality, the undoped gapless system has a critical dimensionality of $d_c=2\alpha$, implying that for $\alpha =1$ (2), $d_c=2$ (4) so that for linear (quadratic) dispersion, the system can only have a disorder-induced semimetal-to-metal transition with the DOS as the order parameter for three (five) dimensions (or above).  For $\alpha=3$, $d_c=6$, and thus the disorder-driven semimetal-to-metal transition considered originally by Fradkin \cite{fradkin1986} becomes trivial
in this case.
Clearly, graphene ($d=2$ and $\alpha=1$) is the marginal case for linearly dispersing systems whereas $d=4$ is the marginal case for parabolic dispersion case ($\alpha=2$).  This makes $\alpha<3/2$ (1) be the interesting nontrivial band dispersion case for 3D (2D) gapless systems.
We refer to clean systems with vanishing (finite) density of states at the Dirac (i.e. band touching) point ($E=0$) as semimetal (metal) throughout this paper, and theoretically investigate the resultant Boltzmann conductivity arising from short-range and long-range impurity scattering.  We are thus asking a question different from (and complementary to) that discussed in Refs.~[\onlinecite{koshino2014,koshino2015,brouwer2014,biswas2014,fradkin1986,goswami2011,roy2014,herbut2014,pixley2015,gurarie2015,roy2015,nand2014}].

We note that the order parameter controlling the actual disorder-driven quantum phase transition \cite{fradkin1986,goswami2011,roy2014,herbut2014,pixley2015}
is the DOS, being zero (finite) in the semimetal (metal) phase, and therefore any connection between this quantum phase transition  and the corresponding behavior of
the charge conductivity, i.e., the semimetal (metal) being defined by vanishing (finite) conductivity, can only be established numerically.  In particular, within the Boltzmann theory, $D(E_F=0)=0$ does not necessarily imply a vanishing conductivity since the transport relaxation time could diverge at $E_F=0$, and conversely, one could have, at least as a matter of principle, a vanishing conductivity for finite values of $D(E_F)$ since the relaxation time could vanish!  One of our goals in the current work is precisely understanding this dichotomy within the Boltzmann theory by investigating various combinations of $d$ and $\alpha$ to see whether it is indeed always true that the DOS and conductivity manifest monolithic behavior with the vanishing (finite) values of $D(E_F=0)$ being necessary and sufficient for vanishing (finite) conductivity in all situations.  We find the answer to this question within the Boltzmann theory to be a definitive negative.

The question of how the Dirac point charge conductivity behaves through the semimetal-metal quantum phase transition has been somewhat controversial \cite{fradkin1986,koshino2014,koshino2015,brouwer2014,nand2014,gurarie2015}. In particular, 
although the conductivity on the metallic side of the transition is always claimed to be finite by virtue of the finite DOS, the conductivity on the semimetal side (with vanishing DOS) has been found to be vanishing (finite) for short-range (long-range) disorder within the self-consistent-Born-approximation (SCBA) for 3D Dirac systems ($d=3$ and $\alpha=1$) in ref.~[\onlinecite{koshino2015}].
The direct numerical calculation\cite{brouwer2014} finds finite conductance at $E=0$ in both phases, but infers that the conductivity actually vanishes (remains finite) in the semimetal (metal). Our earlier Boltzmann theory (using the standard Born approximation for obtaining the relaxation time) gives vanishing (finite) conductivity at the Dirac point for both 2D \cite{hwang2007} and 3D \cite{dassarma2015} Dirac systems using long-range (short-range) disorder in contrast to the SCBA results. Thus, the physics for long-range disorder, which is of considerable experimental relevance because of the invariable presence of random charged impurities in electronic materials, remains open as does the question of whether the interesting semimetal-to-metal transition 
(defined as a transition from vanishing to finite DOS)
at the Dirac point in clean 3D Dirac materials can be distinguished experimentally (at least as a matter of principle) by studying the electrical conductivity through the transition.

The rest of the paper is organized as follows. In section II, which gives the main content of our work, we present our Boltzmann transport theory providing analytical results for the conductivity for various values of $\alpha$ and $d$. We provide a discussion of our results along with a conclusion in section III.
An appendix discussing the temperature dependence of screening completes the presentation.

\section{theory and results}

Although our primary interest is understanding the behavior of the $T=0$ Boltzmann conductivity at the Dirac point for the undoped Dirac material (i.e., the chemical potential at the Dirac point, $k_F=E_F=0$, with no free electrons or holes in the system) as a function of dimensionality ($d$) and band dispersion ($\alpha$) for long-range and short-range disorder, we will consider, for the sake of generality, a system with a finite doping ($k_F,E_F \neq0$) so that our results have general applicability to experimental systems which can never be ideally intrinsic with the chemical potential precisely at the Dirac point. It is easy to figure out the intrinsic Dirac point behavior by taking the appropriate $k_F,E_F \rightarrow 0$ limit of our general density-dependent results. 
We present our analytical results in three different subsections with IIA, B, C giving respectively results for arbitrary dimensions ($d>1$), $d=3$ (and 2), 
and finally (IIC) the results for finite temperatures (with section IIA and B being restricted to $T=0$). We use the word ``Dirac point" to signify the $E=0$ band touching point for all values of $\alpha$ (i.e., arbitrary energy dispersion) although the word Dirac point strictly applies only to the linear chiral band dispersion case ($\alpha=1$).
As mentioned already, our results remain valid for all gapless systems independent of their chirality.


\subsection{Boltzmann Conductivity with  $E=Ak^{\alpha}$ in arbitrary dimension ($d>1$)}

We first consider Dirac materials in arbitrary dimensions ($d>1$).
For a clean system with an energy dispersion
\begin{equation}
E_k = A |\vk|^{\alpha},
\label{eq_ek}
\end{equation}
the density of states in a $d$-dimensional space 
is given by 
\begin{equation}
D_{\alpha}(E) = \frac{1}{2^d\pi^{d/2}}\frac{1}{\Gamma(d/2+1)}\frac{d}{A^{d/\alpha} \alpha} E^{d/\alpha-1},
\label{dos_d}
\end{equation}
where $\Gamma(x)$ is the gamma function,
and the carrier density ($n$), which is defined as particle number per unit $d$-dimensional volume (with only $d=2$, 3 being of physical interest),
is related to the Fermi wave vector ($k_F$) as
\begin{equation}
n = \frac{g}{2^d\pi^{d/2}} \frac{1}{\Gamma(d/2+1)} k_F^d,
\end{equation}
where $g$ is the total degeneracy (i.e., number of fermion flavors including both valley and spin).
Thus, we have $k_F \propto n^{1/d}$.

The Boltzmann charge conductivity can be expressed as
\begin{equation}
\sigma  = \frac{e^2g}{d} \int dE D_{\alpha}(E) v_k^2 \tau_{E_k}
 \left ( - \frac{\partial f_{E_k}}{\partial E_k} \right ),
\label{sigma_d}
\end{equation}
where $v_k = dE_k/dk$ is the velocity of the carrier, and $f_{E_k}$ is the Fermi distribution function.
The scattering time $\tau_{E_k}$ in $d$-dimension (ignoring intervally scattering in the system) is given by 
\begin{eqnarray}
\frac{1}{\tau} & = & \frac{n_i}{\hbar (2\pi)^{d-1}}\frac{E_F^{d/\alpha -1}}{\alpha A^{d/\alpha}} \nonumber \\
&\times &  \int d^{d-1}\Omega |V(q)|^2 f_{\alpha}(\Omega) (1-\cos\theta), 
\label{tau_d}
\end{eqnarray}
where $n_i$ is the $d$-dimensional impurity density, $E_F = A k_F^{\alpha}$ is the Fermi energy,
$d^{d-1}\Omega$ is the $(d-1)$ dimensional solid angle element,  $f_{\alpha}(\Omega)$ is the wavefunction overlap factor arising from the chirality of a system, $\theta$ is the scattering angle between $\vk$ and $\vk'$, and $q = 2k \sin(\theta/2)$. Note that for a nonchiral system $f_{\alpha} (\Omega) = 1$ in Eq.~(\ref{tau_d}).

The Coulomb interaction in the wave vector space is given by the appropriate $d$-dimensional Fourier transform of the Coulomb interaction $V(r)=e^2/\kappa r$, i.e., 
\begin{equation}
V_d(q) = \frac{e^2 (2\sqrt{\pi})^{d-1}\Gamma \left ( \frac{d-1}{2} \right) }{\kappa  q^{d-1}},
\end{equation}
where $\kappa$ is the background lattice dielectric constant of the material.
Note that we take the Coulomb potential to be $1/r$ in all dimensions ($d>1$) where `$r$' is the $d$-dimensional spatial separation.

(1) Screened Coulomb potential -- within the Thomas-Fermi (TF) approximation the screened Coulomb potential can be expressed as
\begin{equation}
V_d(q) = \frac{e^2 (2\sqrt{\pi})^{d-1}\Gamma \left ( \frac{d-1}{2} \right) }{\kappa ( q^{d-1} +  q_s^{d-1})},
\end{equation}
and the screening wave vector $q_s$ in $d$-dimension is given by
($q_s=q_{TF}$ in the TF-approximation)
\begin{equation}
q_s^{d-1} = q_{TF}^{d-1} = \frac{ge^2}{\kappa} (2\sqrt{\pi})^{d-1}\Gamma \left ( \frac{d-1}{2} \right) 
D_{\alpha}(E_F).
\label{qs_d}
\end{equation}
From Eqs.~(\ref{sigma_d}), (\ref{tau_d}), and (\ref{qs_d}) we can calculate the density dependent scattering time and conductivity at zero temperature. In the strong screening limit ($q_s \gg k_{F}$) the scattering time becomes  
\begin{equation}
\tau \propto E_F^{d/\alpha-1} \propto k_F^{d-\alpha } \propto n^{1-\alpha/d},
\label{tau_sigma_1}
\end{equation}
and the Boltzmann conductivity becomes
\begin{equation}
\sigma \propto E_F^{2(d-1)/\alpha} \propto k_F^{2(d-1)} \propto n^{2-2/d}.
\end{equation}

(2) Unscreened Coulomb potential or weak screening limit ($k_F \gg q_s$)--- in this case the scattering time is calculated to be
\begin{equation}
\tau \propto E_F^{(d-2)/\alpha +1} \propto k_F^{d-2 +\alpha} \propto n^{1+(\alpha-2)/d }
\label{tau_sigma_2}
\end{equation}
and the conductivity to be
\begin{equation}
\sigma \propto E_F^{(2d-4)/\alpha + 2} \propto k_F^{2d-4 + 2 \alpha} \propto n^{2+(2\alpha -4)/d}
\end{equation}
 
(3) Density independent screening wave vector, i.e. $q_s =$ constant -- in the strong screening limit $q_s \gg k_F$, we have 
\begin{eqnarray}
\tau & \propto & E_F^{1-d/\alpha} \propto k_F^{\alpha - d} \propto n^{\alpha/d-1} \nonumber \\
\sigma & \propto & E_F^{2-2/\alpha} \propto k_F^{2\alpha -2} \propto n^{2(\alpha-1)/d},
\label{tau_sigma_3}
\end{eqnarray}
and in the weak screening limit ($k_F \gg q_s$) the results are the same as the results for unscreened Coulomb potential.

(4) Short range potential, i.e., $V_q =$ constant -- in this case the results are the same as Eq.~(\ref{tau_sigma_3})
\begin{eqnarray}
\tau & \propto & E_F^{1-d/\alpha} \propto k_F^{\alpha - d} \propto n^{\alpha/d-1} \nonumber \\
\sigma & \propto & E_F^{2-2/\alpha} \propto k_F^{2\alpha -2} \propto n^{2(\alpha-1)/d}.
\label{tau_sigma_4}
\end{eqnarray}
Thus, for $n \rightarrow 0$  the conductivity is only finite for the short range potential and $\alpha =1$ regardless of the system dimensionality. (Note that here $d>1$ throughout.) For all long-ranged disorder, screened or not, the semimetal phase ($n=0$) has vanishing Boltzmann conductivity for all disorder
(except for the density-independent screening, which is exactly equivalent to short-range disorder since $k_F=0$ at the Dirac point).

In the next subsections, we use the general $d$-dimensional analytical results derived above to obtain specific Boltzmann conductivity formula in physical dimensions ($d=2$, 3) for arbitrary band dispersions and for different types (e.g. the long- and short-range disorder potential) of resistive impurity scattering in gapless electron-hole systems.

\subsection{Conductivity in 3D (2D) with arbitrary band dispersion}

In this subsection we obtain the density dependence of the Boltzmann conductivity for the experimentally relevant 3D (and 2D) gapless systems. 
For screened Coulomb disorder we have ($d=3$)
\begin{equation}
\langle V(\vk,\vk') \rangle = \frac{4\pi e^2}{\kappa} \frac{1}{q^2 +
  q_s^2},
\label{vkkp}  
\end{equation}
where 
$q_s$ is the screening wave vector in 3D. 
For $q_s=0$, we get the usual unscreened Coulomb potential in the wave vector space going as $1/q^2$ indicating the long-range $1/r$ behavior of the Coulomb interaction.
From Eq.~(\ref{dos_d}) we have the density of states per electron flavor or species, $D_{\alpha}(E)$, in 3D as (with $E\sim k^{\alpha}$ for the gapless electron-hole band dispersion)
\begin{equation}
D_{\alpha}(E) = \frac{1}{2\pi^2} \frac{E^{3/\alpha -1}}{\alpha A^{3/{\alpha}}}.
\end{equation} 
and within the 3D TF approximation we have for the screening wave vector from Eq.~(\ref{qs_d})
\begin{equation}
q_s^2 = q_{TF}^2 = \frac{4\pi e^2 g}{\kappa}  D_{\alpha}(E_F) \propto E_F^{\frac{3}{\alpha} -1}.
\label{qs2}
\end{equation}
Since $E_F \propto k_F^{\alpha}$ we have
$q_s^2 \propto k_F^{3-\alpha}$,
where $k_F$ is the Fermi wave vector which is related to the carrier density through
$n={g k_F^3}/{6\pi^2}$.

The Boltzmann conductivity at zero temperature is given by 
\begin{equation}
\sigma = e^2 \frac{v_F^2}{3} gD_{\alpha}(E_F) \tau
\label{sigma}
\end{equation}
where $v_F = d E_k/dk \propto k_F^{\alpha-1}$. In the strong screening limit, $q_s \gg k_F$, we have 
\begin{equation}
\tau \propto \frac{q_s^2}{n_i} \propto \frac{k_F^{3-\alpha}}{n_i} \propto \frac{n^{1-\alpha/3}}{n_i},
\end{equation}
Combining with Eq. (\ref{sigma}) we have for three dimensions
\begin{equation}
\sigma \propto \frac{k_F^4}{n_i} \propto \frac{n^{4/3}}{n_i},
\end{equation}
for all $\alpha$. Thus the conductivity vanishes at the Dirac point ($n \rightarrow 0$). This result is universal and is independent of the energy dispersion (i.e., $\alpha$)
and chirality (i.e. the functional form of $f$ in Eq.~(\ref{tau_d})). Thus, the Boltzmann theory 
already predicts zero conductivity at the Dirac point for screened Coulomb disorder, and the semimetal state is preserved independent of disorder strength (i.e., value of $n_i$).
This result applies for arbitrary band dispersion and chirality in gapless electronic systems at the Dirac point.

For unscreened Coulomb potential or very weak screened potential ($k_F \gg q_s$) we have 
\begin{equation}
\tau \propto \frac{k_F^{1+\alpha}}{n_i} \propto \frac{n^{(1+\alpha)/3}}{n_i},
\end{equation}
and
\begin{equation}
\sigma \propto \frac{k_F^{2\alpha + 2}}{n_i} \propto \frac{n^{2(\alpha + 1)/3}}{n_i}.
\end{equation}
Thus, for unscreened disorder the conductivity is a non universal function of density (i.e., dependent on $\alpha$), but it nevertheless vanishes in the $n\rightarrow 0$ semimetal limit for all values of $\alpha$ (and arbitrary chirality).

For short range disorder $V(q)$ is a constant and we have
\begin{equation}
\tau \propto \frac{k_F^{\alpha-3}}{n_i} \propto \frac{n^{\alpha/3 -1}}{n_i},
\end{equation}
giving
\begin{equation}
\sigma \propto \frac{k_F^{2\alpha -2}}{n_i} \propto \frac{n^{2(\alpha-1)/3}}{n_i}.
\label{sig_3D_zero}
\end{equation}
In this case the conductivity becomes finite for only $\alpha = 1$ as $n \rightarrow 0$, and the semimetal Dirac point manifests a finite conductivity in the Boltzmann theory
for all values  of the disorder strength (and chirality).

For a 2D system we can calculate the density dependent conductivity also, obtaining \\
(1) strong screening Coulomb disorder ($q_s \gg k_F$) 
\begin{equation}
\sigma \propto \frac{k_F^2}{n_i} \propto n,
\end{equation}
(2) unscreened or weak screening Coulomb disorder ($q_s \ll k_F$)
\begin{equation}
\sigma \propto \frac{k_F^{2\alpha}}{n_i} \propto \frac{n^{\alpha}}{n_i},
\end{equation}
(3) short-range disorder
\begin{equation}
\sigma \propto \frac{k_F^{2\alpha-2}}{n_i} \propto \frac{n^{\alpha-1}}{n_i}.
\end{equation}
Again, for linear Dirac dispersion ($\alpha=1$), the 2D Dirac point conductivity is finite for the semimetal
for short-range disorder only, vanishing for screened or unscreened disorder.
These 2D results also apply for all gapless systems independent of their chirality.

\subsection{Temperature dependent conductivity arising from energy averaging in the semimetal}

Before considering the temperature dependent conductivity we first calculate the activated carrier density at finite temperatures. For a gapless intrinsic $d$-dimensional system, in which conduction and valence bands meet at the Dirac point and therefore the chemical potential is located at the Dirac point ($\mu=0$), the activated electron density is given by
\begin{equation}
n(T) = g \int \frac{d^dk}{(2\pi)^d} \frac{1}{e^{\beta E} + 1},
\end{equation}
where $g$ is the total degeneracy and $\beta = 1/k_BT$. We can rewrite this equation as
\begin{eqnarray}
n(T) & = &\frac{g}{2^d\pi^{d/2}}\frac{1}{\Gamma(d/2+1)}\frac{d}{A^{d/\alpha} \alpha} \int_0^{\infty} dE\frac{E^{d/\alpha-1}}{e^{\beta E} + 1} \nonumber \\
&=& C (k_BT)^{d/\alpha}, 
\end{eqnarray}
where $C$ is a constant. Thus the activated electron density is proportional to $T^{d/\alpha}$.
In 2D (3D) the thermally activated density increases linearly with temperature for quadratic (cubic) dispersion $E=Ak^2$ ($E= Ak^3$)
whereas for linear band dispersion, the activated density increases as $T^2$ ($T^3$) for 2D (3D) systems.

The conductivity Eq.~(\ref{sigma_d}) can be written as
\begin{equation}
\sigma = B \int_0^{\infty}dE E^{\frac{d-2}{\alpha}+1} \tau(E) \left ( - \frac{\partial f_E}{\partial E} \right ),
\end{equation}
where $B$ is a constant.
Assuming that $\tau \propto E^{\beta}$ and is independent of the temperature, the energy averaging of the scattering time gives the following temperature dependent Boltzmann conductivity in the $n\rightarrow 0$ limit (i.e., intrinsic undoped systems)  
\begin{equation}
\sigma \propto T^{\frac{d-2}{\alpha} + 1+ \beta}.
\label{eq_36}
\end{equation}
The energy dependent scattering time is given in Eqs.~(\ref{tau_sigma_1}), (\ref{tau_sigma_2}), (\ref{tau_sigma_3}), and (\ref{tau_sigma_4}) for different situations, and substituting the corresponding $\beta$ in Eq.~(\ref{eq_36}) one obtains the temperature dependent Boltzmann conductivity for the semimetal (i.e., $n=0$) for different types of disorder.

Writing the temperature dependent Boltzmann conductivity in the semimetal (i.e. $E_F=0$) as 
\begin{equation}
\sigma(T) \sim T^{\gamma},
\label{eq_37}
\end{equation}
we get 
\begin{equation}
\gamma = \frac{d-2}{\alpha} + 1 + \beta,
\label{eq_38}
\end{equation}
where $\beta$ is the exponent defining the energy dependence of the transport relaxation time ($\tau \sim E^{\beta}$). We note that although Eqs.~(\ref{eq_37}) and (\ref{eq_38}) give the main temperature dependence of the Boltzmann conductivity, we have ignored the (presumably weak) T-dependence arising from screening (see Appendix A) as we assume $q_s$ to be temperature independent in deriving Eq.~(\ref{eq_38}).

For specificity, we note that in 3D linear dispersion Dirac semimetal ($d=3$ and $\alpha=1$) we get $\gamma=4$, 4, 0 for screened Coulomb, unscreened Coulomb, and short-range disorder, respectively (the corresponding $\gamma=2$, 3, 1 for 3D chiral quadratic dispersion, $d=3$ and $\alpha=2$). Thus, the temperature dependent semimetallic ($E_F=0$) conductivity of the gapless system, in principle, provides direct information about both the energy dispersion (and hence the noninteracting DOS) and the disorder potential (i.e., long-range or short-range). We note here also that the scaling of the Boltzmann conductivity, $\sigma(E_F)$ at $T=0$, on the Fermi energy $E_F$, given by $\sigma \sim E_F^{\delta}$ with the exponent $\delta$ (as given above in Eqs.~(\ref{tau_sigma_1}) -- (\ref{tau_sigma_4})), follows exactly the same scaling law as thermal scaling, i.e., $\delta=\gamma$ as will be discussed further later in section III.
(The equality $\delta=\gamma$ holds quite generally up to logarithmic accuracy provided that screening itself is weakly temperature-dependent and the main temperature dependence arises from the thermal averaging.)

In particular, let us consider $d=\alpha=3$ whence the 3D DOS is finite in the semimetal phase (instead of vanishing at the Dirac point, $E_F=0$, as in the linear dispersion situation). For $d=\alpha=3$, we have $\gamma=4/3$, 8/3, 4/3 for screened Coulomb, unscreened Coulomb, and
short-range disorder, respectively. 
This implies that, although the DOS itself is finite for $d=\alpha=3$ case, the corresponding Dirac point Boltzmann conductivity {\it always} vanishes for all disorder types keeping the system a semimetal (as far as the conductivity goes) independent of the fact the DOS itself is finite! By contrast, $d=3$ and $\alpha=1$ case has $\gamma= 4$ (screened Coulomb), 4 (unscreened Coulomb), 0 (short-ranged), implying that the Dirac point Boltzmann conductivity vanishes for Coulomb disorder (screened or unscreened), but is finite for short-ranged disorder although the DOS at $E=0$ vanishes in this case.  Thus, we actually get the intuitively unexpected result that in 3D clean systems, in the presence of short-range disorder, the `semimetallic' state with vanishing DOS at $E=0$ (i.e. the $\alpha=1$, $d=3$ case) has finite conductivity whereas the `metallic' state with finite DOS at $E=0$ (i.e., $\alpha=3$, $d=3$ case) has vanishing conductivity!  Thus, at least for the clean system, where the DOS can be tuned by the dispersion exponent $\alpha$, there is no connection between the DOS and the Boltzmann conductivity.
Given that the $d=3$, $\alpha=1$ case has vanishing DOS at $E_F=0$ whereas the $d=3$, $\alpha=3$ case has finite DOS at $E_F=0$, we conclude that the temperature dependent intrinsic semimetallic conductivity (i.e., the actual value of the exponent $\gamma$)
could, in principle, distinguish the situation of finite DOS from the vanishing DOS in a gapless system, but not whether the conductivity itself is finite or not.
All the temperature-dependent results derived in this sub-section apply both to chiral and non-chiral gapless systems.

\section{discussion and conclusion}

Our main theoretical finding is that the semiclassical Boltzmann conductivity always vanishes at the semimetallic ``Dirac point" (i.e., $E_F=0$) for all clean gapless systems in all dimensionalities for all Coulomb disorder (screened or unscreened) and short-ranged disorder except for the short-range uncorrelated white-noise disorder in the linear dispersion case ($\alpha=1$) which provides finite dc conductivity in all dimensions for all disorder strength. We have obtained general analytical results for the conductivity, $\sigma(E_F,T,\alpha,d;n_i)$, as a function of Fermi energy (or equivalently carrier density), temperature, band dispersion ($E \sim k^{\alpha}$), system dimensionality, and disorder strength or type 
with $E_F=0$ denoting the ``semimetallic" intrinsic phase and $E_F\neq 0$ being the generic ``metallic" extrinsic phase with free carriers. In the presence of (screened or unscreened) Coulomb disorder and short-ranged disorder with $\alpha \neq 1$, we find $\sigma(T=0;E_F=0)=0$ for all values of $d$ as long as $n_i \neq 0$ whereas for short-range disorder in the linear dispersion situation we find $\sigma(T=0;E_F=0) \neq 0$ for all $d$ and $\alpha=1$. Our qualitative results with respect to conductivity (i.e., whether $\sigma$ is zero or finite) thus cannot distinguish between zero of finite DOS phases in the 3D Dirac system (with $\alpha=1$) at $E_F=0$ since both situations (vanishing or finite DOS at $E=E_F=0$) will give either zero conductivity (Coulomb disorder) or finite conductivity (short-range disorder) at $T=0$. Thus, the Boltzmann transport theory does not directly manifest any direct or indirect effect of the semimetal-to-metal quantum phase transition at the Dirac point.
It is, however, worth while to mention that the pure Dirac linear dispersion case ($\alpha=1$) is special (at least with respect to short-range disorder) since it is the only
situation which allows for a finite dc conductivity (in both $d=2$ and 3) at $E_F=0$ in spite of the DOS being zero.
If we use Boltzmann conductivity for $E_F=0$ as the `order parameter' (rather than the DOS itself), we conclude that the disorder-driven critical transition discussed in 
Refs.~[\onlinecite{koshino2014,koshino2015,brouwer2014,biswas2014,fradkin1986,goswami2011,roy2014,herbut2014,pixley2015,gurarie2015,roy2015,nand2014}]
should not be called a semimetal-metal transition simply because the DOS changes from being zero below to being finite above the transition since the expression `metal' typically implies a finite $T=0$ conductivity, and there is no evidence whatsoever that this is true in the putative `metallic' phase for long-range disorder as $T$ and $E_F$ go to zero. 
A recent direct numerical work\cite{brouwer2014} investigating the conductance for disordered 3D Dirac systems explicitly considered short-range disorder.

Our detailed analytical scaling results for $\sigma(E_F,T)$, however, suggest some qualitative possibilities for distinguishing between semimetallic ($D=0$ at $E_F$) and metallic ($D \neq 0$ at $E_F$) phases through the power law dependence of the conductivity on $E_F$ (or equivalently density $n$) and/or temperature $T$. We find (see section II for details)
\begin{equation}
\sigma(T,E_F=0) \sim T^{\gamma},
\end{equation}
and
\begin{equation}
\sigma(E_F,T=0) \sim E_F^{\delta},
\end{equation}
where the scaling exponents $\gamma$ and $\delta$ depend on $\alpha$, $d$, and disorder. While $\gamma$ is defined in Eqs.~(\ref{eq_37}) and (\ref{eq_38}), we provide $\delta$ below for various situations (by collecting together the results from section II)
\begin{equation}
\delta = \left\{ 
  \begin{array}{l l}
    2(d-1)/\alpha & \quad \text{screened Coulomb disorder } \\
    \frac{2(d-2)}{\alpha} + 2 & \quad \text{unscreened Coulomb disorder} \\
    2 - 2/\alpha,  & \quad \text{short-range disorder}    
  \end{array} \right. .
\label{eq_39}
\end{equation}
We note that for density-independent screening (i.e., assuming that the screening wave vector $q_s$ is independent of $k_F$ and is a constant), the strongly screened Coulomb disorder gives the same results as the short-ranged disorder. Comparing Eqs.~(\ref{eq_38}) and (\ref{eq_39}), and putting in appropriate $\beta$ values, we conclude $\delta = \gamma$, which is an important verification of our theory since we expect this exponent identity based on general scaling arguments.

We note that the scaling exponents $\gamma=\delta$ in the Boltzmann theory do in general depend on both $\alpha$ and $d$ (as well as on $\beta$, with $\tau \sim E^{\beta}$, which is determined by the type of disorder) except for the pure short-ranged disorder which is independent of dimensionality. Thus, we have the following identity valid for all disorder (provided the temperature dependence of screening is negligible and the temperature dependence in the conductivity is arising entirely from energy averaging in the Boltzmann theory)
\begin{equation} 
\gamma = \delta.
\end{equation}
The ``semimetallic" behavior in conductivity manifests itself for $T>T_F=E_F/k_B$ with $\sigma \sim T^{\gamma}$ and the ``metallic" behavior manifests itself for $T<T_F$ with $\sigma \sim E_F^{\delta}$, and $\delta = \gamma$. This physics is schematically depicted in Fig.~1 which also serves to summarize our theoretical results showing that the ``high-temperature" (i.e. $T>T_F$ which is also the low-density) regime corresponds to the semimetal 
and the ``low-temperature" (i.e., $T<T_F$, low density) regime corresponds to the ``metallic" regime with a finite $T=0$ value of conductivity.

\begin{figure}[t]
	\centering
	\includegraphics[width=.8\columnwidth]{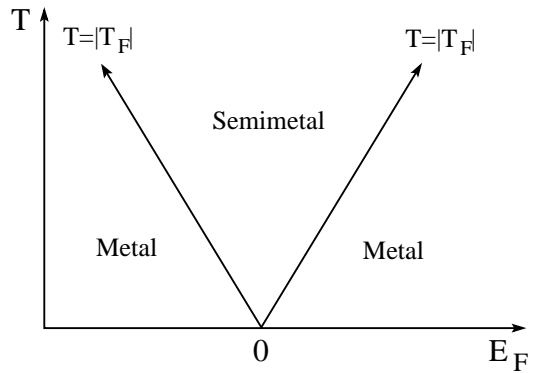}
	\caption{
The figure shows the metal-semimetal generic phase diagram defined by the Boltzmann conductivity with the high-temperature ($T>|T_F|$) regime manifesting the semimetallic scaling $\sigma \sim T^{\gamma}$ whereas the low-temperature ($T<|T_F|$) regime manifesting the metallic scaling $\sigma \sim E_F^{\delta}$ (with $\gamma=\delta$ in the simple theory). All conductivities except for linear dispersion with short-range disorder gives vanishing conductivity at the Dirac point at $T=0$.
The explicit analytical values of the scaling exponents $\gamma=\delta$ are given in the text.
}
\label{fig1}
\end{figure}

We note that the Boltzmann theory predicts $\sigma(T=0,E_F=0)=0$ even for $\alpha=3$ (2) in $d=3$ (2) dimensions for {\it all} disorder models (including short-range disorder) although the corresponding 3D (for $E\sim k^3$) and 2D (for $E\sim k^2$) DOS is finite at $E_F=0$. 
Thus, within the Boltzmann transport theory at least,
having finite DOS at $E_F=0$ is {\it not} synonymous with a finite metallic conductivity at $E_F=0$, and one should be careful in automatically identifying metals (with finite conductivity) and semimetals (with zero conductivity) with having finite or zero DOS respectively!

We note that to the extent exact numerical work \cite{brouwer2014} and SCBA \cite{koshino2014} find the disorder-driven semimetal-to-metal quantum phase transition to be directly connected with the Dirac point conductivity increasing from zero to a finite value at a finite disorder (also as shown in Ref. \onlinecite{fradkin1986}), the disorder model used in these theories  is short-ranged disorder which provides a nongeneric answer even in the Boltzmann theory of ours where in $d=2$ and 3 we find the linearly dispersing Dirac system to always have a finite conductivity even in the $T=0$, $E_F=0$ limit in the presence of only short-range disorder. For long-range disorder, the conductivity vanishes for $E_F=0$, and indeed the SCBA theory also no longer gives results consistent with the putative quantum phase transition.\cite{koshino2015}
The issue of quantum criticality associated with long-range disorder remains an open question at this stage.

Our work indicates that the semimetal-metal transition, even if it is theoretically allowed, will be very difficult, if not impossible, to observe experimentally. Any real 3D linearly dispersing ($\alpha=1$)
Dirac system will obviously have some Coulomb disorder because of the inevitable presence of charged impurities. At low enough $E_F$ (i.e. low enough doping), this Coulomb disorder is {\it always} relevant for transport since it leads to zero conductivity (i.e., infinite resistivity) whereas any short-range disorder (even if it is important at high carrier density for large $E_F$) becomes completely irrelevant at the Dirac point (as $E_F \rightarrow 0$) since it gives rise to finite conductivity. This implies that transport properties at the Dirac point ($E_F=0$) will be completely determined by Coulomb disorder leading to zero conductivity at the Dirac point (except for any finite-temperature effects we discuss above). This remains true, within the Boltzmann theory, independent of whether the DOS is finite or not at the Dirac point, and therefore, the Dirac point conductivity is always likely to be semimetallic in nature (see Fig. 1) in the presence of any long-range disorder.

A brief discussion of graphene (i.e., $d=2$, $\alpha=1$) transport properties may be germane in this context \cite{dassarma2011,peres} since $d=2$ is the critical dimensionality for the semimetal-to-metal transition (for $\alpha=1$) implying that infinitesimal disorder produces finite DOS at the Dirac point in graphene, thus essentially making all undoped graphene samples everywhere being ``metallic" 
with a universal Dirac point ($E_F=0$)
intrinsic conductivity of $\sigma_D = 4e^2/\pi h$. Such a universal conductivity has never been measured in graphene dc transport experiments where the transport measurements typically agree with \cite{hwang2007,tan2007}
the Boltzmann theory predictions for long-range Coulomb disorder with $\sigma \sim E_F \propto n$ for higher carrier density with the conductivity crossing over at low carrier density ($E_F \sim 0$) to a broad nonuniversal minimum, which depends weakly on the amount of disorder (and other details), arising from inhomogeneous electron-hole puddle formation in the system which has been studied extensively. \cite{adam2007,li2012,chen2008,martin2008,hwang2007,tan2007}
Our experience with graphene hints at what is likely to be seen in 3D Dirac transport when detailed experimental data become available. For $E_F \gg k_B T$, we expect that the conductivity will follow the Boltzmann prediction \cite{dassarma2015} going as $E_F^{\delta}$, 
and then depending on the details (e.g., temperature, amount of disorder, etc.), at some low values of $E_F$, there will be a crossover to the intrinsic $\sigma \sim T^{\gamma}$ ``semimetallic" behavior which will persist as long as $k_B T > E_F$. Eventually, 
at some disorder-dependent
low density, there will be a crossover to the puddle physics with the conductivity showing a nonuniversal broad minima around the Dirac point similar to what is extensively seen in graphene. This puddle physics is in some sense reminiscent of the physics of ``rare regions" discussed in Ref.~[\onlinecite{nand2014}] although the details are completely different since the puddles discussed by us arise from the density inhomogenity inherent in Coulomb disorder in the absence of screening by carriers which must happen around $E_F \sim 0$. \cite{efros} The conclusion, however, is the same that even the ``semimetallic" phase at $E_F=0$ always has a finite local density of states as the system nucleates electron and hole puddles (with fluctuating local density, but vanishing average density for $E_F=0$) in order to screen the impurity induced Coulomb potential. These finite local density of states at the Dirac point have directly been observed experimentally in graphene through imaging microscopy 
techniques, \cite{martin2008,puddle} and we believe that they must exist in 3D Dirac systems also at low enough carrier doping \cite{skinner2014}.

One important caveat to note here is that all our transport results are obtained using the Boltzmann transport theory with the disorder-induced transport relaxation time calculated in the leading order Born approximation.  Strictly speaking such a theory is, in principle, inapplicable at the Dirac point with vanishing carrier density (where $k_F$ vanishes) since the dimensionless parameter $k_Fl>1$, where $l$ is the mean free path, must apply for the theory to be meaningful.  In fact, this is a limitation not just for the Born approximation and the Boltzmann theory, but for all perturbative theoretical approaches to calculating the charge conductivity at the Dirac point since $k_Fl$ is no longer a control parameter at the Dirac point.  This is why our theory is carried out at finite carrier density (i.e., finite $k_Fl$) and the Dirac point is approached as a limiting case of $k_F$ going to zero.  As long as there is no singularity at the Dirac point, such a limiting procedure for obtaining the Dirac point conductivity should be at least qualitatively valid although our theory is strictly valid only for a low (but nonzero) carrier density so that $k_Fl$ remains finite.  We mention here that simple improvements of the theory, for example, using the self-consistent Born approximation or including the higher-order Born scattering terms is also not justified since all of these approximations are expansions in $1/k_Fl$, which becomes ill-defined at the Dirac point since $k_F=0$ there!  In particular, we do not believe that the self-consistent Born approximation, which is sometimes used, is a better approximation at all since it is completely uncontrolled (and thus, its domain of validity unknown) and includes only selective higher-order impurity scattering diagrams leaving out many terms in each order in an {\it ad hoc} manner.  
One great advantage of the Born approximation-based Boltzmann theory is that it is known to be valid and accurate at higher carrier density (and has had great success in describing carrier transport properties in semiconductors, metals, and graphene\cite{dassarma2011}), and thus, unless there is a quantum phase transition with decreasing carrier density, the theory should remain qualitatively valid as the Dirac point is approached by decreasing $k_F$ albeit with increasing quantitative inaccuracy.  At this stage, we do not see a way to develop a better transport theory precisely at the Dirac point (i.e. $k_F=0$), which in any case is a set of measure zero not particularly relevant for experiments (e.g., any finite temperature immediately relaxes the zero-density condition, making the Boltzmann theory applicable in a formal sense).  As we show in our work (see Fig.~1), the important physics is that the intrinsic (extrinsic) semimetallic (metallic) behavior manifests itself in the charge conductivity for $T>(<)$ $T_F$ in gapless systems, and the strict zero-density or zero-temperature limit is unnecessary.

We emphasize that the transport theory used in the current work is the Boltzmann theory where the effect of disorder is included only in calculating the transport scattering rate (and not in the DOS itself), and our goal is to correlate the Boltzmann transport (not critical transport) behavior with the non-interacting DOS in the clean system. This approach is known to work very well in graphene\cite{adam2007,li2012,martin2008,hwang2007,tan2007}, where the Boltzmann theory based calculated conductivity\cite{adam2007,li2012,hwang2007} agrees well with experimental results \cite{chen2008,martin2008,tan2007} for systems with arbitrary disorder in spite of graphene being the critical case (satisfying $d=2\alpha$).  
If quantum critical properties become important in an experimental situation (and we know of no direct experimental evidence for the predicted quantum critical behavior $d_c=2\alpha$ in any physical system), then our Boltzmann theory is restricted to weakly disordered systems for 
$d \leq 2\alpha$, and to systems with disorder strength less than the critical value for $d>2\alpha$.  But, our main goal of connecting the Boltzmann conductivity with the clean system DOS for a gapless system is completely independent of these disorder-driven critical properties considerations.

We conclude by saying that we have analytically obtained the scaling dependence of the Boltzmann conductivity $\sigma(E_F,T)$ in gapless systems with arbitrary band dispersion and dimensionality due to scattering by Coulomb and short-range disorder. Our results 
are generically valid for both chiral and non-chiral gapless systems, and
illuminate the behavior of semiclassical conductivity in the semimetallic (zero density of states) and the metallic (finite density of states) gapless systems (when the chemical potential is near the band touching or ``Dirac" points), showing that the power law dependence (neglecting logarithmic corrections)
of the disorder-limited conductivity could, in principle, distinguish the semimetallic behavior ($\sigma \sim T^{\gamma}$) at high temperatures ($T>T_F$) from the metallic behavior ($\sigma \sim E_F^{\delta}$) at low temperatures.

\section*{acknowledgments} 

This work is supported by LPS-CMTC (SDS and EHH), JQI-NSF-PFC (SDS), and 
National Research Foundation of Korea (NRF-2014R1A2A2A01006776) (EHH).

\appendix

\section{Temperature dependent screening in the semimetal}

In deriving the main results of this paper (at finite temperatures), we neglected any temperature dependence arising from the screening itself, concentrating instead on the universal temperature dependence of the conductivity arising from the energy averaging which dominates the high-temperature conductivity for $T>T_F$ in the semimetallic phase (where the carrier density and hence the Fermi energy are small by definition).  In principle, however, screening itself could have temperature dependence which can be quite subtle and rather important for non-chiral systems, particularly in 2D systems \cite{dassarma1986} 
with parabolic dispersion where the $2k_F$ `kink' in the zero-temperature polarizability leads to nonanalytic corrections of $O(T/T_F)$ in the Boltzmann conductivity arising from screened Coulomb disorder \cite{dassarma2004}.  In chiral systems of interest in the current work, $2k_F$ back-scattering is suppressed, and therefore, the temperature dependence of conductivity is much weaker for screened Coulomb disorder \cite{dassarma2015,hwang2009}
Nevertheless, the screening wave vector $q_s$ may itself pick up a temperature dependence which we consider below (this effect is neglected in the main body of our work, but is straightforward to include \cite{dassarma2015,hwang2009}).

At first, it may seem that since $q_s$ is basically the density of states at the Fermi level, it cannot have any temperature dependence since the DOS, by definition, is temperature-independent.  This is, however, untrue since at finite temperature, even for $E_F=0$ at $T=0$, an averaging over energy must be done in order to calculate any physical quantity.  This averaging over the Fermi distribution function leads to a T-dependence in screening even in the semimetallic phase at the Dirac point (and $q_s$ is no longer just simply the DOS at $E_F$).

The temperature dependent TF screening wave vector can be calculated from the d-dimensional polarizability at $q=0$,
\begin{equation}
\Pi(q=0,T) = g \int dE D_{\alpha}(E) \left ( - \frac{\partial f_E}{\partial E} \right ).
\end{equation}
Then, we have the $q_s$ as
\begin{equation}
q_s(T) \propto T^{\frac{d-\alpha}{\alpha (d-1)}}.
\end{equation}
For the linear dispersion ($\alpha=1$) we find that $q_s$ increases linearly with temperature regardless of dimensionality.
We note that,  for $d>\alpha$, which is true for 2D and 3D Dirac systems, screening at the Dirac point is enhanced (i.e. $q_s$ increases) with increasing temperature, and thus leads to insulating or
semimetallic temperature dependence (i.e. an increasing conductivity with increasing temperature) which is the same behavior as that from energy averaging studied in the main part of the paper.  For finite doping with $E_F>k_B T$, the temperature dependence of $q_s$ is exponentially suppressed by $\exp(-T/T_F)$ and is no longer important.
Also, disorder itself may suppress the temperature-dependence of screening by modifying the polarizability function as has been considered in the literature for parabolic systems \cite{dassarma1986,R1,R3}.


\begin{thebibliography}{999}

\bibitem{dassarma2011}S. Das Sarma, S. Adam, E. H. Hwang, and E. Rossi,
Rev. Mod. Phys. {\bf 83}, 407 (2011). 

\bibitem{peres} N. M. R. Peres, Rev. Mod. Phys. {\bf 82}, 2673 (2010).


\bibitem{geim2004} K. S. Novoselov, A. K. Geim, S. V. Morozov, D. Jiang, Y. Zhang, S. V. Dubonos, I. V. Grigorieva, and A. A. Firsov, Science {\bf 306}, 666 (2004).

\bibitem{zyuzin2012} A. A. Burkov, M. D. Hook, and L. Balents, Phys. Rev. B {\bf 84}, 235126 (2011);
A. A. Burkov and L. Balents, Phys. Rev. Lett. {\bf 107}, 127205 (2011); 
X. Wan, A. M. Turner, A. Vishwanath, and S. Y. Savrasov, Phys. Rev. B {\bf 83}, 205101 (2011);
P. Hosur, S. A. Parameswaran, and A. Vishwanath
Phys. Rev. Lett. {\bf 108}, 046602 (2012).

\bibitem{dassarma2015} S. Das Sarma, E. H. Hwang, and Hongki Min,
Phys. Rev. B {\bf 91}, 035201 (2015).

\bibitem{skinner2014} B. Skinner, Phys. Rev. B {\bf 90}, 060202(R) (2014);
N. Ramakrishnan, M. Milletari, and S. Adam, arXiv:1501.03815 (2015)

\bibitem{koshino2014}Y. Ominato and M. Koshino, Phys. Rev. B {\bf 89}, 054202 (2014).

\bibitem{koshino2015} Y. Ominato and M. Koshino, arXiv:1410.0155.
 

\bibitem{brouwer2014} B. Sbierski, G. Pohl, E. J. Bergholtz, and P. W. Brouwer, Phys. Rev. Lett. {\bf 113}, 026602 (2014).

\bibitem{biswas2014} R. R. Biswas and S. Ryu, Phys. Rev. B {\bf 89}, 014205 (2014).

\bibitem{fradkin1986} E. Fradkin, Phys. Rev. B {\bf 33}, 3263 (1986).

\bibitem{goswami2011} P. Goswami and S. Chakravarty, Phys. Rev. Lett. {\bf 107}, 196803 (2011).


\bibitem{roy2014} B. Roy and S. Das Sarma, Phys. Rev. B {\bf 90}, 241112(R)  (2014).

\bibitem{herbut2014} K. Kobayashi, T. Ohtsuki, K.-I. Imura, and I. F. Herbut, Phys. Rev. Lett. {\bf 112}, 016402 (2014).

\bibitem{pixley2015} J. Pixley, P. Goswami, and S. Das Sarma, arXiv:1502.07778.

\bibitem{gurarie2015} S. V. Syzranov, V. Gurarie, and L. Radzihovsky, arXiv:1411.4635 (2014);
S.V. Syzranov, L. Radzihovsky, and V. Gurarie, arXiv:1402.3737.

\bibitem{roy2015} B. Roy and P. Goswami, unpublished.


\bibitem{nand2014} R. Nandkishore, D. A. Huse, and S. L. Sondhi, Phys. Rev. B {\bf 89}, 245110 (2014).

\bibitem{adam2007} S. Adam, E. H. Hwang, V. Galitski, and S. Das
  Sarma, Proc. Natl Acad. Sci. {\bf 104}, 18392 (2007).


\bibitem{li2012} E. Rossi and S. Das Sarma, Phys. Rev. Lett. {\bf 101}, 166803 (2008); Q. Li, E. Rossi, and S. Das Sarma, Phys. Rev. B {\bf 86}, 235443 (2012); E. H. Hwang and S. Das Sarma, Phys. Rev. B {\bf 82}, 081409 (2010); Q. Li, E. H. Hwang, and S. Das Sarma, Phys. Rev. B {\bf 84}, 115442 (2011).

\bibitem{chen2008} J.-H. Chen, C. Jang, S. Adam,
  M. S. Fuhrer, E. D. Williams, and M. Ishigami, Nat. Phys. {\bf 4},
  377 (2008). 

\bibitem{martin2008} J. Martin, N. Akerman, G. Ulbricht, T. Lohmann, J. H. Smet, K. V. Klitzing, and A. Yacoby, Nat. Phys. {\bf 4}, 144 (2008).

\bibitem{aleiner2006}I. L. Aleiner and K. B. Efetov, Phys. Rev. Lett. 97, 236801 (2006).

\bibitem{hwang2007} E. H. Hwang, S. Adam, and S. Das Sarma,
Phys. Rev. Lett. {\bf 98}, 186806 (2007).

\bibitem{tan2007} Y.-W. Tan, Y. Zhang, K. Bolotin, Y. Zhao, S. Adam,
  E. H. Hwang, S. Das Sarma, H. L. St\"{o}rmer, and P. Kim, 
Phys. Rev. Lett. {\bf 99}, 246803 (2007); J. Heo, H. J. Chung, S. H. Lee, H. Yang, D. H. Seo, J. K. Shin, U. I. Chung, S. Seo, E. H. Hwang, and S. Das Sarma, Phys. Rev. B {\bf 84}, 035421 (2011).

\bibitem{efros} B. I. Shklovskii and A. L. Efros, {\it Electronic Properties of Doped
Semiconductors} (Springer-Verlag, Berlin, 1984). 

\bibitem{puddle} J. Xue, J. Sanchez-Yamagishi, D. Bulmash, P. Jacquod, A. Deshpande, K. Watanabe, T. Taniguchi, P. Jarillo-Herrero, and B. J. LeRoy, Nature Mat. {\bf 10}, 282 (2011).

\bibitem{dassarma1986} S. Das Sarma, Phys. Rev. B {\bf 33}, 5401 (1986).

\bibitem{dassarma2004} S. Das Sarma and E. H. Hwang, Phys. Rev. B {\bf 69}, 195305 (2004);
Solid State Commun. {\bf 135}, 579 (2005).

\bibitem{hwang2009} E. H. Hwang and S. Das Sarma, Phys. Rev. B {\bf 79}, 165404 (2009).

\bibitem{R1}  P. G. de Gennes, J. Phys. Radium {\bf 23}, 630 (1962).
\bibitem{R3} S. Das Sarma and W. Y. Lai, Phys. Rev. B {\bf 32}, 1401 (1985). 

\end{thebibliography}
\end{document}